\documentstyle[epsfig]{elsart}

\begin{document}
\begin{frontmatter}
\title{Mean Field Theory for a driven Granular Gas of Frictional Particles}
\author{Raffaele Cafiero\thanksref{mail1}}
\address{P.M.M.H., Ecole Sup\`erieure de Physique et de Chimie 
Industrielles (ESPCI), 10, rue Vaquelin-75251 Paris cedex 05, FRANCE}
\author{Stefan Luding}
\address{Institute for Computer Applications 1, 
Pfaffenwaldring 27, 70569 Stuttgart, GERMANY}
\thanks[mail1]{cafiero@pmmh.espci.fr}
\begin{abstract}
We propose a mean field (MF) theory for a homogeneously driven 
granular gas of inelastic particles with Coulomb friction. 
The model contains three parameters, a normal restitution 
coefficient $r_n$, a maximum tangential restitution coefficient 
$r_t^m$, and a Coulomb friction coefficient $\mu$. The parameters
can be tuned to explore a wide range of physical situations. 
In particular, the model contains the frequently used 
$\mu \rightarrow \infty$ limit as a special case.
The MF theory is compared with the numerical simulations 
of a randomly driven monolayer of spheres for a wide range of parameter 
values. If the system is far away from the clustering instability 
($r_n \approx 1$), we obtain a good agreement between mean 
field and simulations for $\mu=0.5$ and $r_t^m=0.4$, but 
for much smaller values of $r_n$ the agreement is less good. We 
discuss the reasons of this discrepancy and 
possible refinements of our computational scheme.
\end{abstract}
\end{frontmatter}

\section{Introduction}
\vskip -0.5cm
Granular gases \cite{herrmann98} are usually described as collections 
of macroscopic particles with rough surfaces and dissipative interactions. 
In order to study them, kinetic theories 
\cite{jenkins85b,huthmann97,noije98} and
numerical simulations \cite{luding95b} were applied
for special boundary conditions. The dynamics of the system is assumed to be
dominated by two-particle collisions, modeled by their
asymptotic states: A collision is characterized by the velocities
before and after the contact, and the contact is assumed to be 
instantaneous. In the simplest model, one describes inelastic
collisions by normal restitution $r_n$ only. However, 
surface roughness is important \cite{huthmann97,luding95b},
since it allows for an exchange of translational and rotational energy.
Here we briefly sketch a study of a model where a Coulomb friction 
law with coefficient $\mu$ and a tangential restitution 
coefficient $r_t$ account for tangential inelasticity and friction 
\cite{luding95b,walton93}.
We first introduce the model, then we describe a MF theory for the 
driven granular gas of rough spheres, and finally, 
we compare analytical results and numerical simulations. In 
the conclusions we discuss possible future refinements 
of the computational scheme.

\section{The Model}
\vskip -0.5cm
We consider
$N$ $3$-dimensional spheres of mass $m$ and diameter $2a$ 
interacting via a hard-core potential, confined on a $2$-dimensional (2D)
layer of linear size $L$, with periodic boundary conditions. Inelasticity
and roughness are described by a normal restitution $r_n$, 
a Coulomb friction law with friction $\mu$, and a
tangential restitution $r_t$ which depends on $r_n$, $\mu$ 
and the collision angle $\gamma_c$ for sliding contacts and on a 
maximum tangential restitution $r_t^m$ for sticking contacts, when the 
tangential elasticity becomes important.
When two particles $1$ and $2$ collide, their velocities after collision 
depend on the velocities before collision through a collision 
matrix whose elements depend on $r_n$, $\mu$, $\gamma_c$, and $r_t^m$.
Thus we calculate the momentum change using a model that is consistent
with experimental measurements \cite{walton93}.
From the momentum conservation laws for linear and angular direction, 
energy conservation, and Coulomb's law we get the change of 
linear momentum of particle $1$ as a
function of $r_n$, $\mu$, and $r_t$ \cite{luding95b}.
The change of the normal component of the relative velocity 
depends on the normal restitution $r_n$, which is 
a tunable parameter, while 
the change of the tangential component of the relative velocity  
depends on by the tangential restitution coefficient 
$r_t= \min \left [ r_t^C, r_t^m \right ]$, where $r_t^m$ is the 
coefficient of maximum tangential restitution, 
$-1\leq \! r_t^m \! \leq \! 1$.The quantity  
$r_t^C$ is determined using Coulomb's law
such that for solid spheres $r_t^C = - 1 - (7/2) \mu (1+r_n) 
\cot \gamma_c$ with the collision angle $\pi/2 < \gamma_c \le \pi$ 
\cite{luding95b}. Here, we simplified the tangential
contacts in the sense that exclusively Coulomb-type interactions, i.e. $%
\Delta P^{(t)}$ is limited by $\mu \Delta P^{(n)}$, or sticking contacts
with the maximum tangential restitution $r_t^m$ are allowed \cite{luding95b}. 

\section{The mean field Theory}
\vskip -0.5cm
We start from the results of Huthmann and Zippelius \cite{huthmann97} 
for a freely cooling gas of infinitely 
rough particles. Here we apply their MF theory to the case 
of a driven gas of rough particles, which is the most 
common experimental situation \cite{experiments}. 
The MF kinetic theory of Huthmann and Zippelius 
 is formulated for a gas of rough particles with {\em constant} 
tangential restitution $r_t$, corresponding to the limit 
$\mu\!=\!\infty$ in our model. It is based on a pseudo--Liouville--operator 
formalism and on the assumption of a homogeneous state, 
with a Gaussian probability distribution of 
translational and rotational energies. 
The main outcome of this approach is a set of coupled evolution 
equations for the translational and rotational temperatures 
$T_{tr}$ and $T_{rot}$ \cite{huthmann97}. Here, we write down and 
solve the MF equations for $T_{tr}$ and $T_{rot}$ for a granular gas 
of rough particles in which the translational velocities 
are subjected to a random uncorrelated Gaussian driving of variance 
$\xi_0^2$. These equations read for a 2D layer of spheres
\begin{equation}
\begin{array}{ll}
  \frac{d}{dt}  T_{tr}(t) = \frac{2}{D}\left[
  - G A T_{tr}^{3/2} + G B T_{tr}^{1/2} T_{rot}\right] +m \xi_0^2 \\ 
  \frac{d}{dt}  T_{rot}(t) = \frac{2}{2D-3}\left[
  G B T_{tr}^{3/2} - G C T_ {tr}^{1/2} T_{rot}\right]\\
G  = 4 a n \sqrt{\frac{\pi}{m}} \chi\;,\;  
A  =\frac{1-r_n^2}{4}+\frac{\eta}{2}(1-\eta)  \\
B  = \frac{\eta^2}{2q}\;,\;
C  = \frac{\eta}{2q} \left (1-\frac{\eta}{q} \right )~,
\end{array}
\label{mfrp}
\end{equation}
\vspace{-0.6cm}~\\
where $\eta\! =\! {q(1+r_t)}/{(2q+2)}$, $q\!=\!2/5$ 
for spheres, $n$ is the gas 
density and $\chi$ is the pair correlation function at contact. \\
We can use the Verlet-Levesque \cite{verlet82} 
approximation in 2D $\chi\!=\!(1-7\phi/16)/(1-\phi)^2$ 
where $\phi$ is the volume fraction of the gas.
For long times the system
 approaches a steady state. By imposing $\frac{d}{dt}  T^{eq}_{tr}\!=\!0, 
\frac{d}{dt} T^{eq}_{rot}\!=\!0$ we get the equilibrium temperatures
\begin{equation}
T_{tr}^{eq}= m \left(\frac{\xi_0^2 \sqrt{\pi}}
             {2 \gamma \Omega_D \chi  n a^{D-1}}\right)^{2/3}
~~{\rm and }~~
\frac{T_{rot}^{eq}}{T_{tr}^{eq}}=R=\frac{2(1+r_t)}{9-5r_t}\,\,,\,\,
\end{equation}
\vspace{-0.6cm}~\\
where $\Omega_D=2 \pi^{D/2} / \Gamma (D/2) =2 \pi$ for $D=2$ and
$\gamma=\frac{1-r_n^2}{4}+\frac{1}{49}(1+r_t)\left(6-r_t \right)-
{(5/49)(1+r_t)^3}/{(9-5r_t)}$ for spherical particles.
By linearizing the set of Eqs.\ (\ref{mfrp}) around $T_{tr}^{eq}$ 
and $T_{rot}^{eq}$ we get the final approach to the steady state: 
$\delta T_{rot}(t) \simeq  R \delta T_{tr}(t)$ and 
$T_{tr}(t)\!-\!T_{tr}^{eq}\!=\!\delta 
T_{tr}(t)\simeq \delta T_{tr}(0) \exp[-3 \gamma \omega t]$.
The quantity $\omega=\Omega_D\chi n a^{D-1}\sqrt{\frac{T^{eq}_{tr}}{\pi m}}$ 
is the Enskog collision frequency for elastic particles 
at the temperature $T^{eq}_{tr}$, and 
$t_c=(3 \gamma \omega)^{-1}\propto \gamma^{-2/3}$ is a 
characteristic relaxation time \cite{noije98}.
The MF Eqs.\ (\ref{mfrp}) can be applied to the 
three parameter model in the limit $\mu=\infty$,
see Fig.\ \ref{fig1}.\\
However, experimental measurements are well reproduced by this model
only for finite $\mu$.
Here we investigate the possibility to describe the  effect
of finite friction $\mu$ by replacing $r_t$ in the Eqs.\
(\ref{mfrp}) with its average $\langle{r_t}\rangle$
over the probability density
$P[\sin(\gamma_t)]$ of the normalized impact parameter
$b/(2a)=\sin( \gamma_t)$, with the condition
$r_t \leq r_t^m$.  The used simplifying approach ($\gamma_c \approx 
\gamma_t$) ignores the fluctuations of $r_t$ due to its dependence
on the rotational degree of freedom, and we
expect that it is correct only in some trivial limiting cases.
Nevertheless, this simple approximation allows us to
realize that for $r_t^m=0.4, \mu=0.5, r_n\approx0.9$, corresponding
to many experiments, simulations fit well with the modified MF theory,
as we will se below. If the molecular
chaos hypothesis is valid $P[\sin(\gamma_t)]=1$, and we get
$\langle{r_t}\rangle\!=\!-\!1\!+\!\frac{7}{2} \mu (1+r_n)\!\!
~\ln\!\left(\!{\sqrt{1+c^2}}\!+\!{c}\!\right)$,
with $c\!=\!{\frac{2}{7}(1+r_t^m)}/({\mu(1+r_n)})$.
\section{The simulations}
\begin{figure}[b]
\begin{center}
\vspace{0.3cm}~\\
\epsfig{file=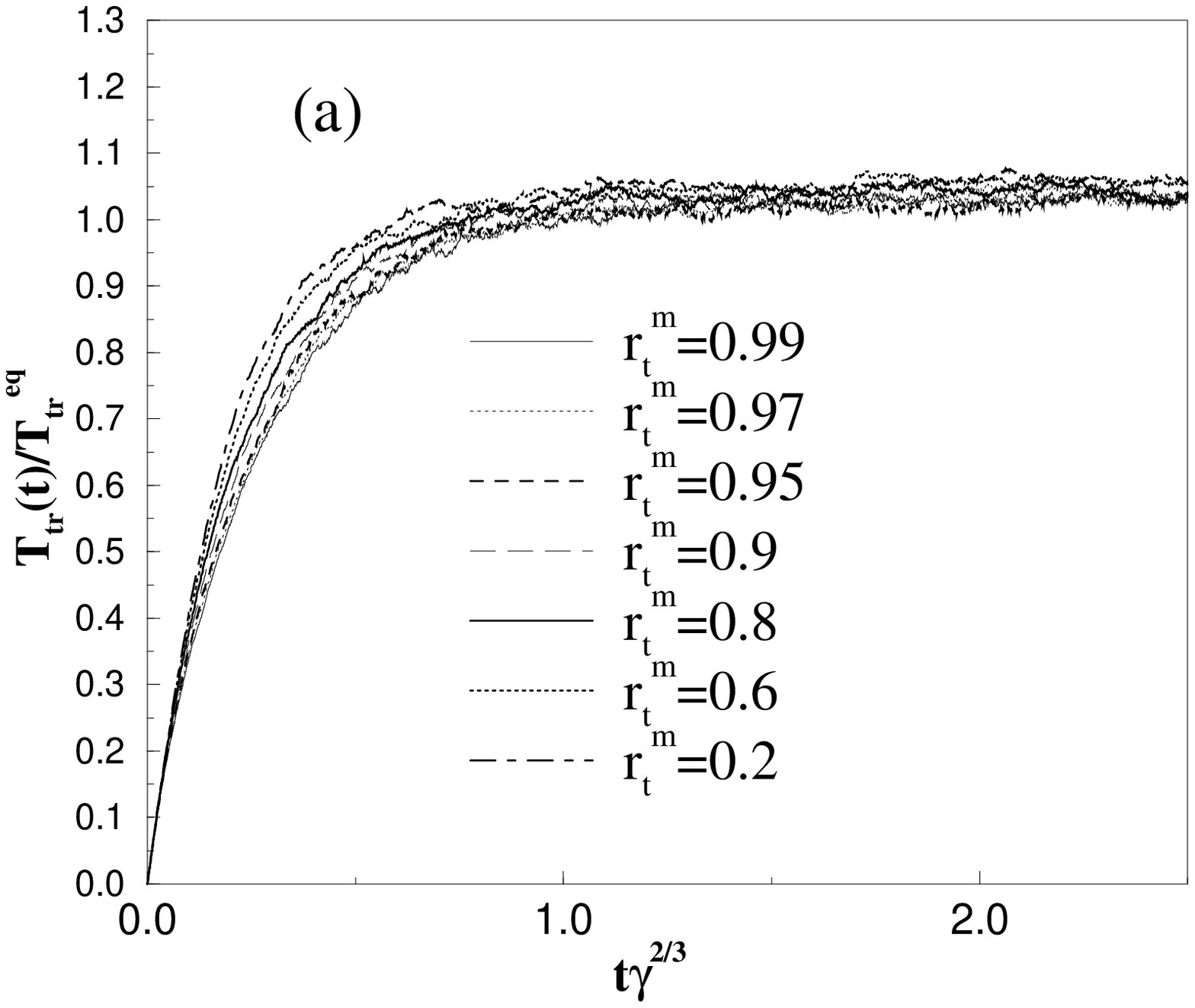,height=4.5cm,width=5.7cm} \hspace{0.5cm}
\epsfig{file=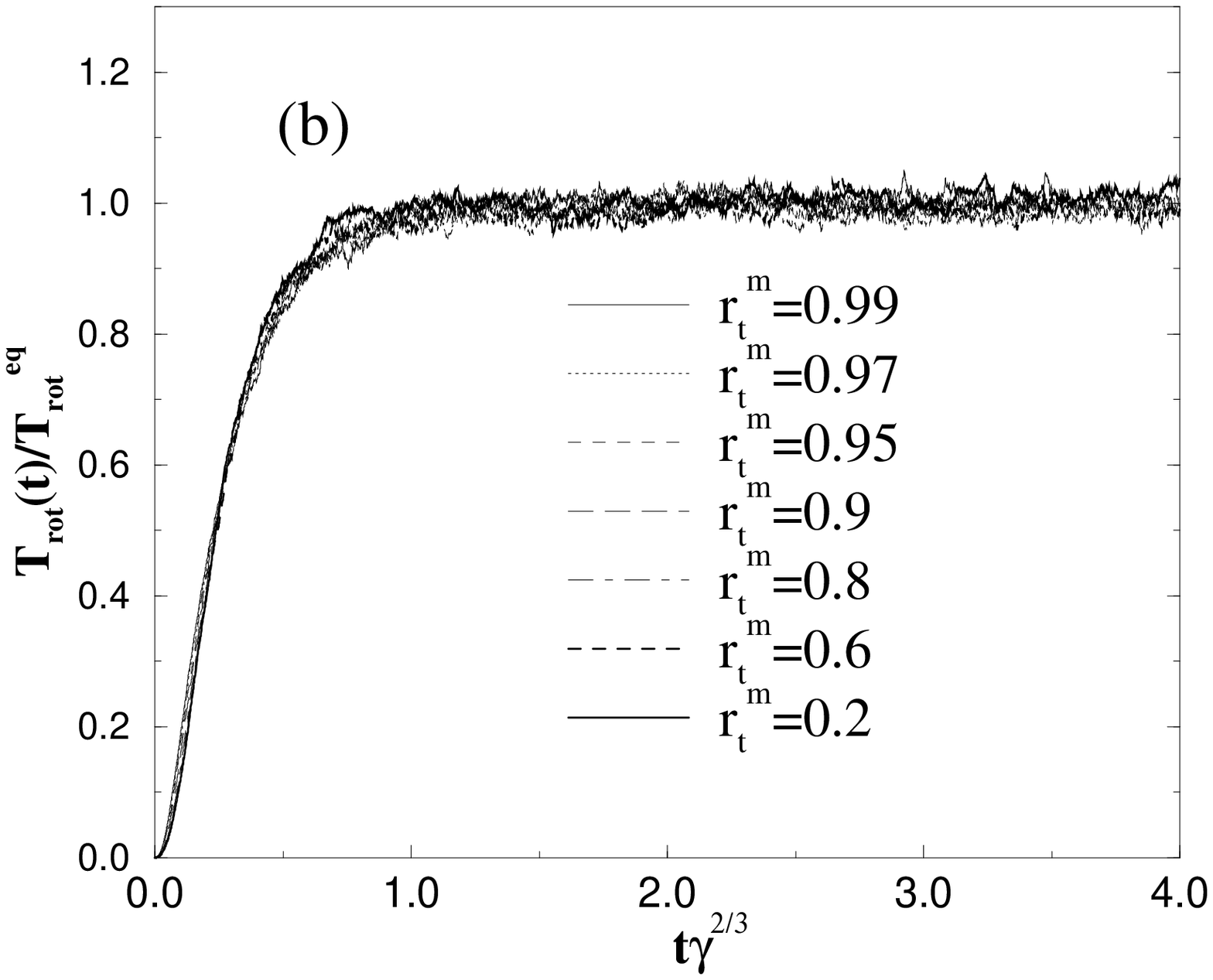,height=4.5cm,width=5.7cm} \\
\epsfig{file=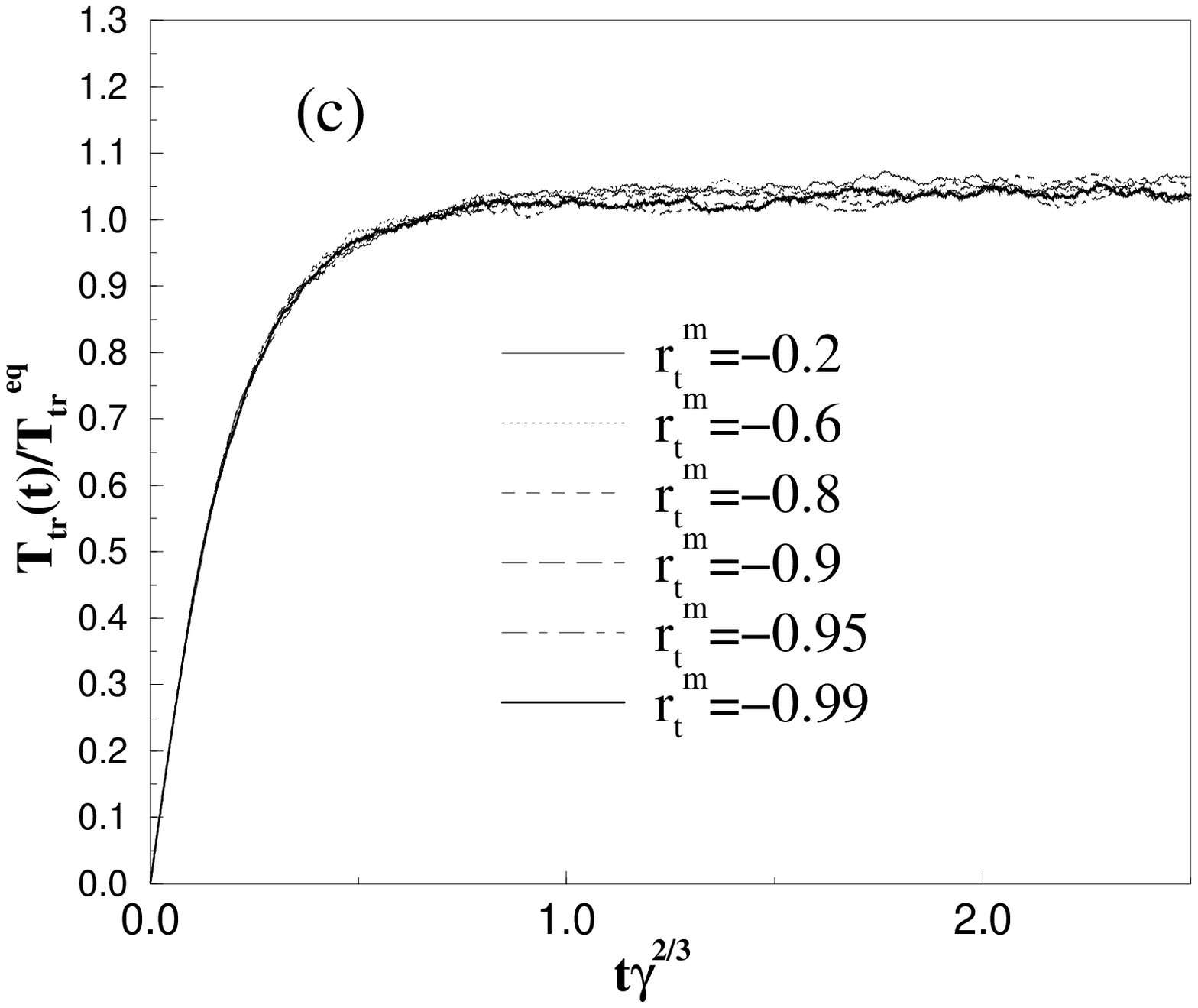,height=4.5cm,width=5.7cm} \hspace{0.5cm}
\epsfig{file=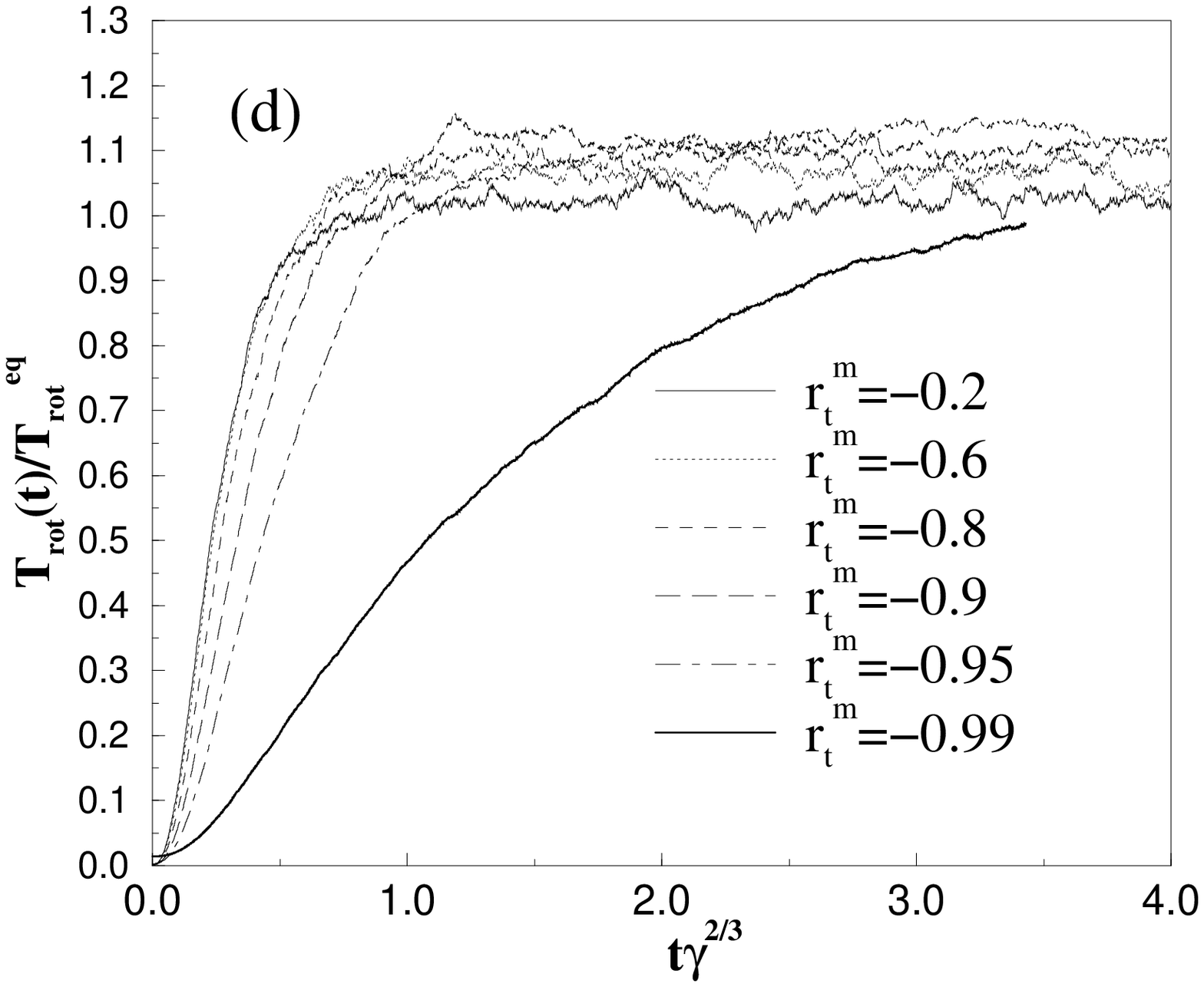,height=4.5cm,width=5.7cm}
\end{center}
\vspace{-0.3cm}
\caption{\footnotesize{(a)-(b) Rescaled translational and rotational
temperatures $T_{tr}(t)/T_{tr}^{eq}$, $T_{rot}(t)/T_{rot}^{eq}$
versus the rescaled time $t \gamma^{2/3}$, for $r_n=0.95$, $\mu=10^7$
and positive $r_t^m$; (c)-(d) the same as (a)-(b) but for negative 
$r_t^m$.}}
\label{fig1}
\vspace{0.1cm}
\end{figure}

\vskip -0.5cm
Here we compare numerical simulations of a
randomly driven monolayer of spheres, performed by using an
Event Driven (ED) algorithm \cite{luding95b}, with the MF predictions
(for details see \cite{noije98,luding95b,mazighi94mcnamara94,cafiero99}).
Every simulation is equilibrated without driving with $r_n=1$ and
$r_t^m=-1$. Then inelasticity and driving are switched on.
We used a fixed volume fraction
$\phi=0.34$, $N=11025$ and different values of
$r_n$, $r_t^m$, $\mu$. In Fig.\ \ref{fig1}a-d, the
translational and rotational temperatures for fixed
$r_n=0.95$, $\mu=10^7$ and
different values of $r_t^m$ are rescaled
with the MF equilibrium temperatures and plotted versus
the rescaled time $t/t_c\propto t \gamma^{2/3}$. The high value of
$\mu$ allows to decouple normal and
tangential momentum ($r_t=r_t^m$).
The agreement is good, even for low positive values of $r_t^m$.
\begin{figure}
\hfill
\begin{center}
\epsfig{file=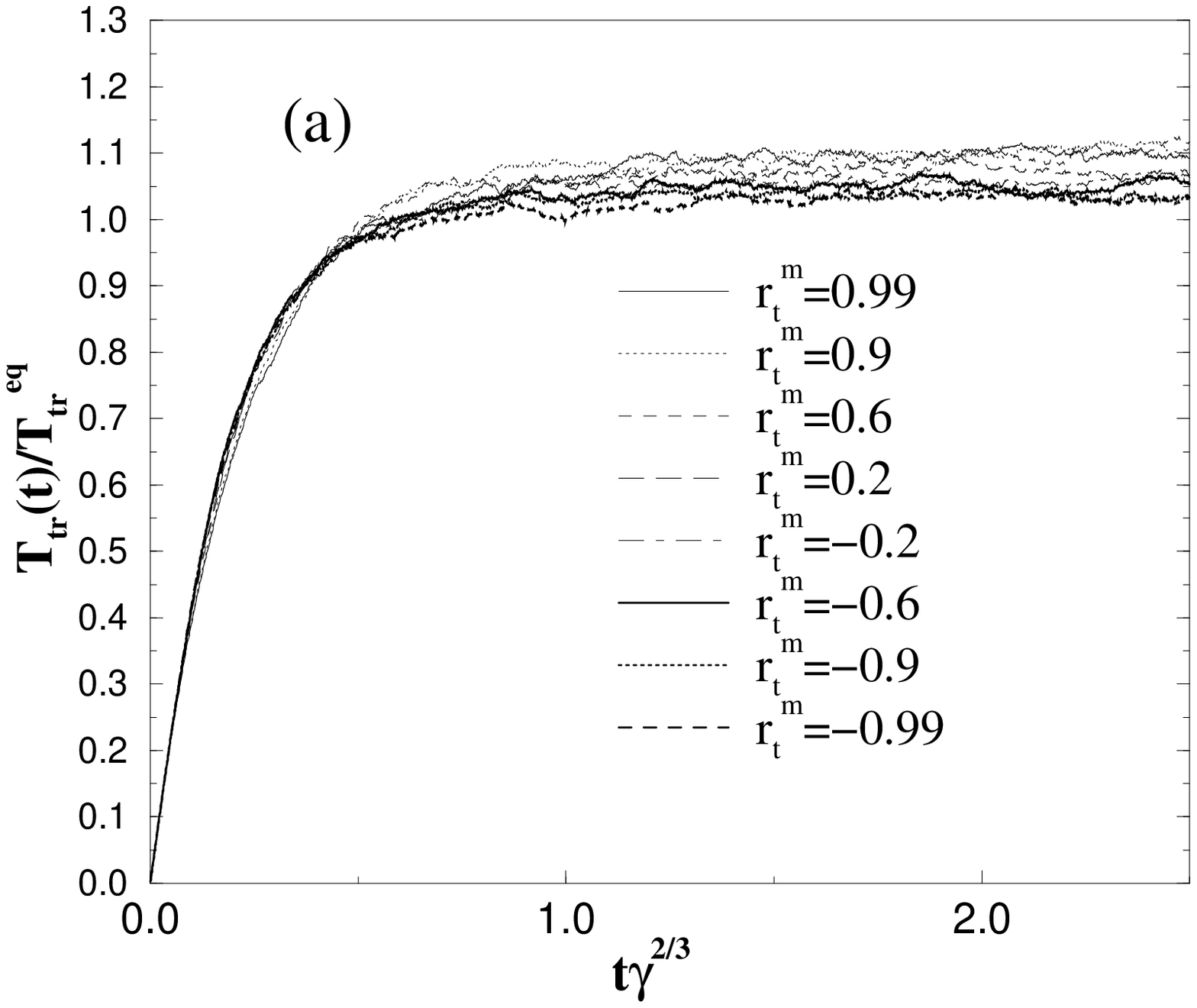,height=4.5cm,width=5.7cm} \hspace{0.5cm}
\epsfig{file=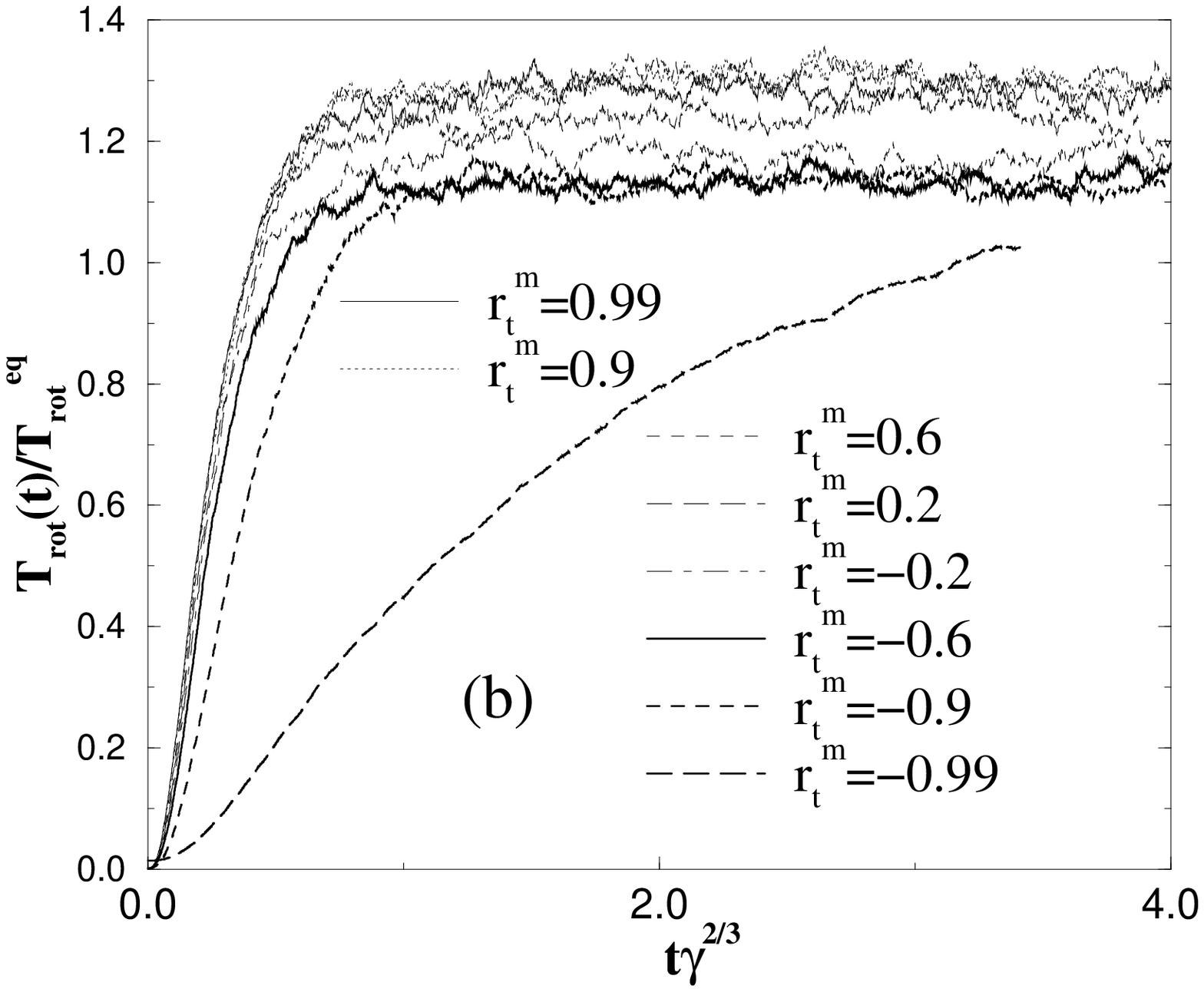,height=4.5cm,width=5.7cm}
\hfill
\end{center}
\vspace{-0.1cm}
\caption{\footnotesize{(a)-(b) Rescaled translational and rotational
temperatures $T_{tr}(t)/T_{tr}^{eq}$, $T_{rot}(t)/T_{rot}^{eq}$
versus the rescaled time $t\gamma^{2/3}$, for $r_n=0.95$, $\mu=0.5$.}}
\label{fig2}
\vspace{-0.1cm}
\end{figure}
\begin{figure}
\begin{center}
\vspace{0.1cm}~\\
\epsfig{file=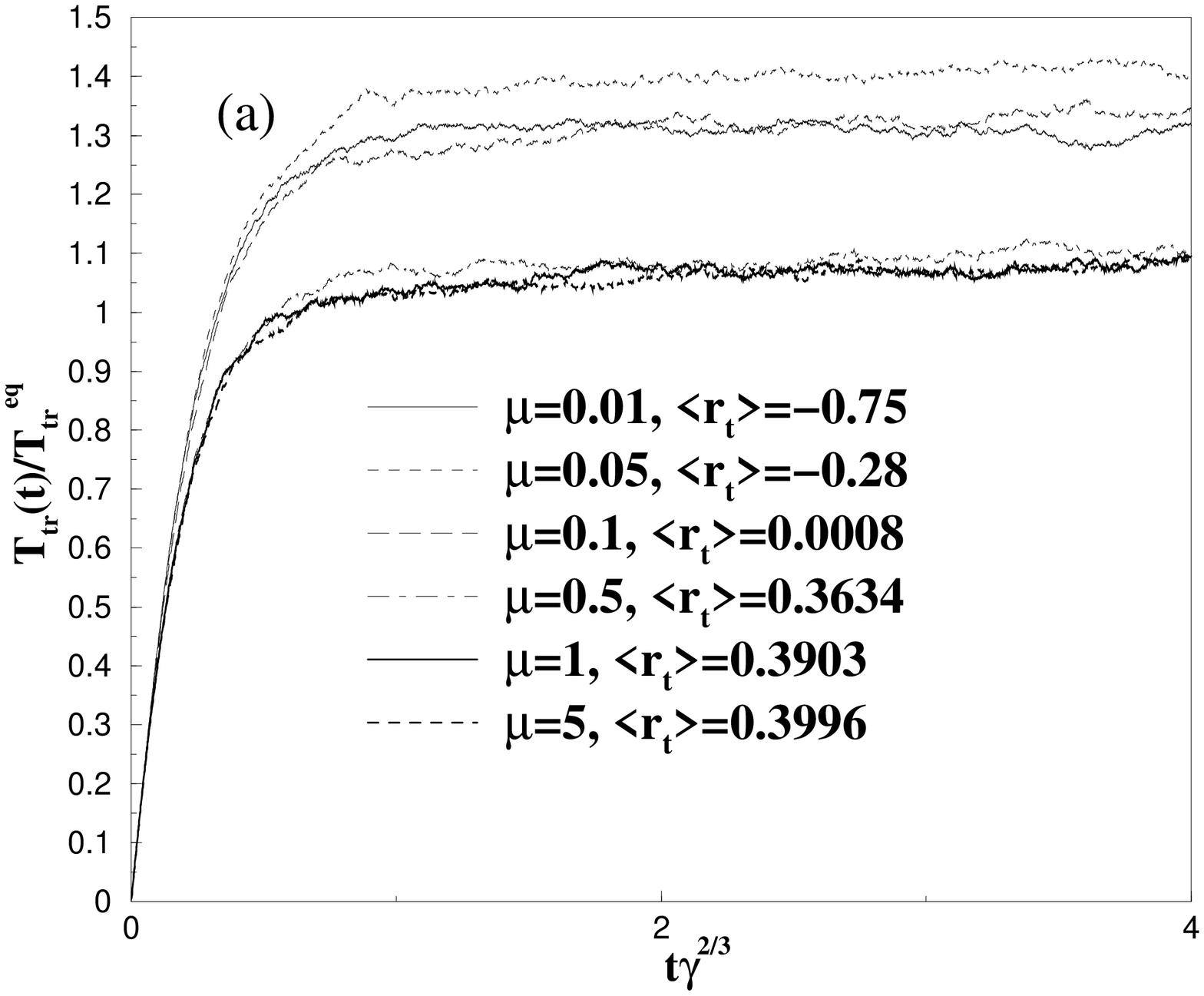,height=5.7cm,width=4.5cm,angle=-90} \hspace{0.5cm}
\epsfig{file=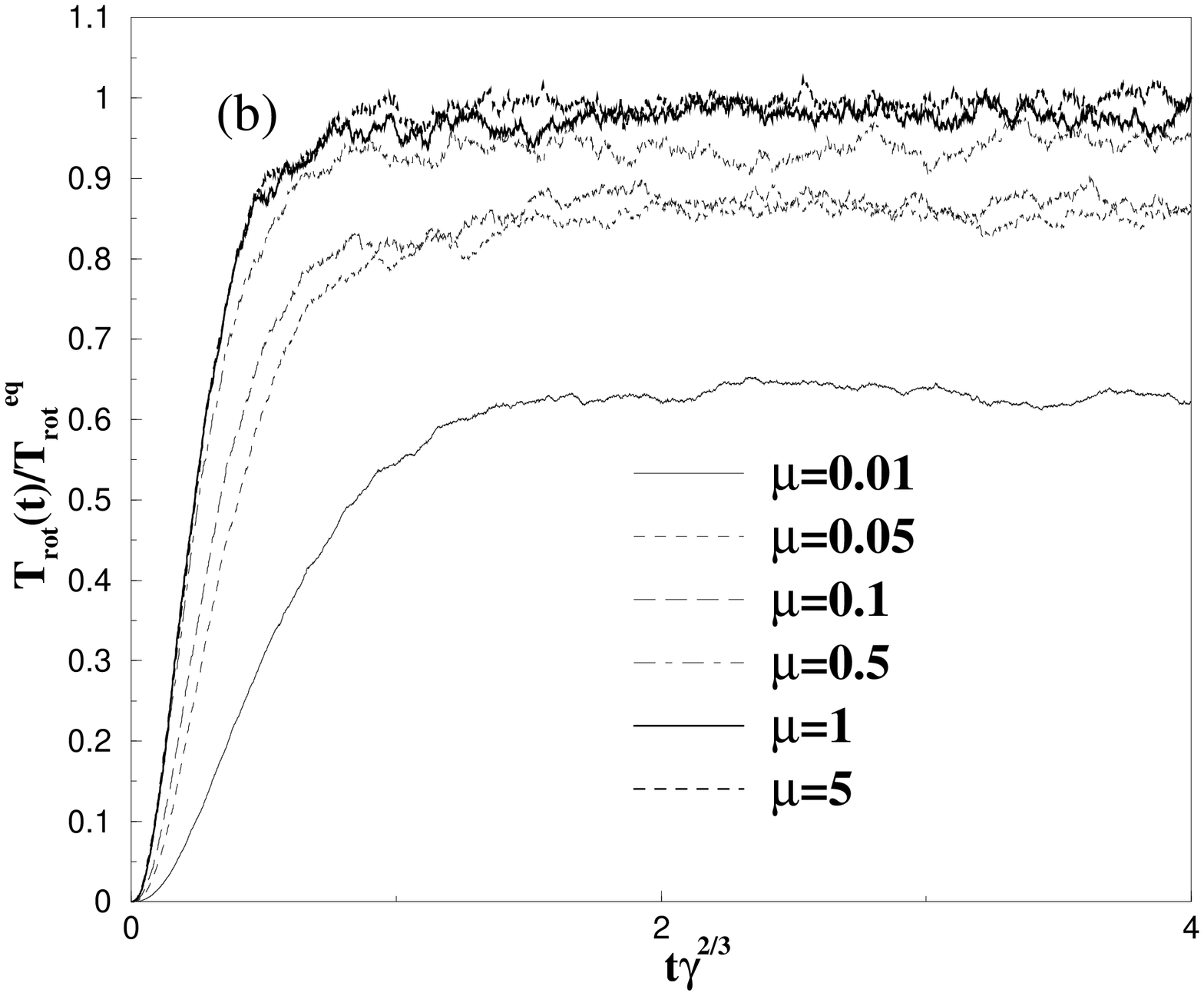,height=5.7cm,width=4.5cm,angle=-90}\\
\epsfig{file=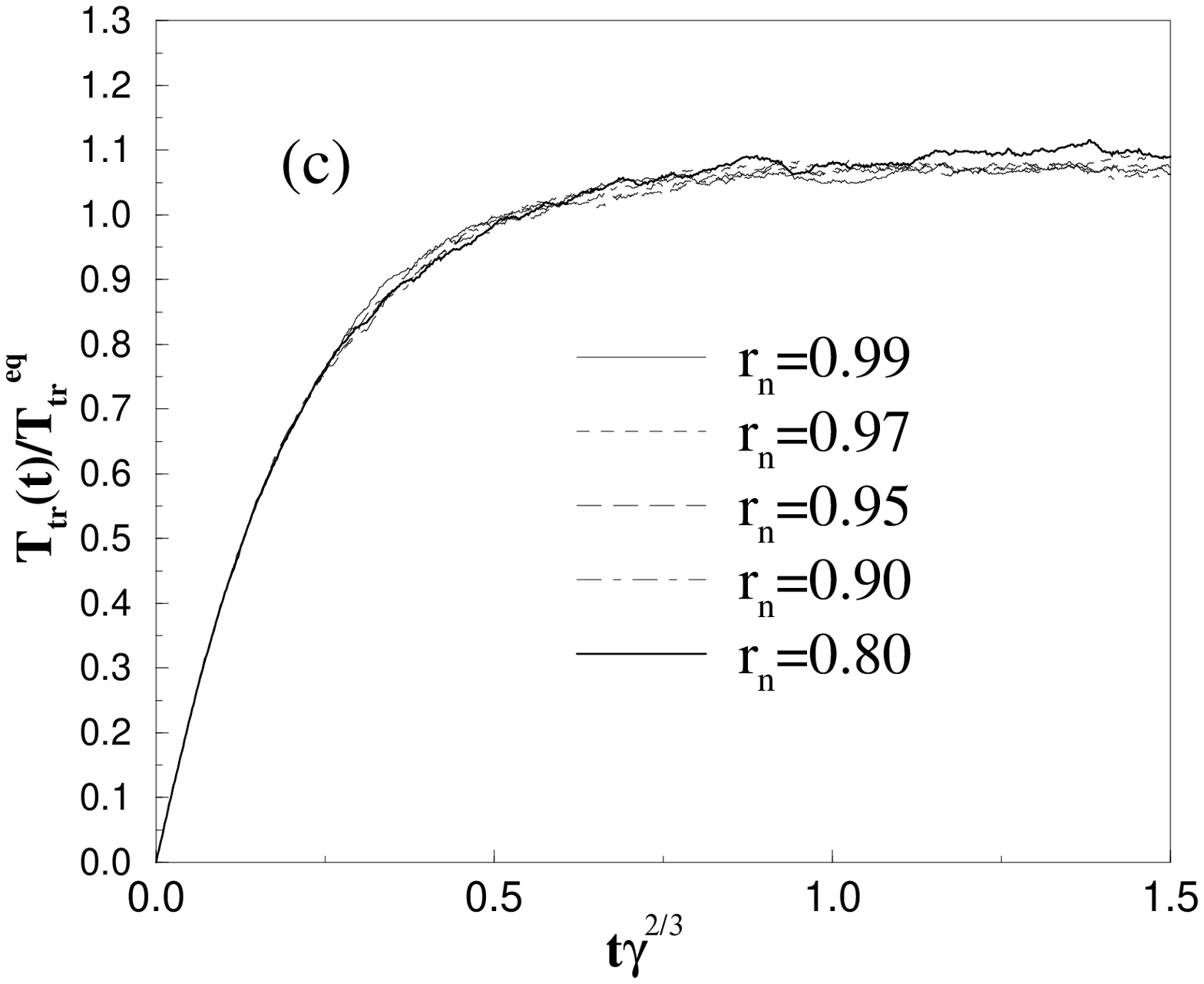,height=4.5cm,width=5.7cm} \hspace{0.5cm}
\epsfig{file=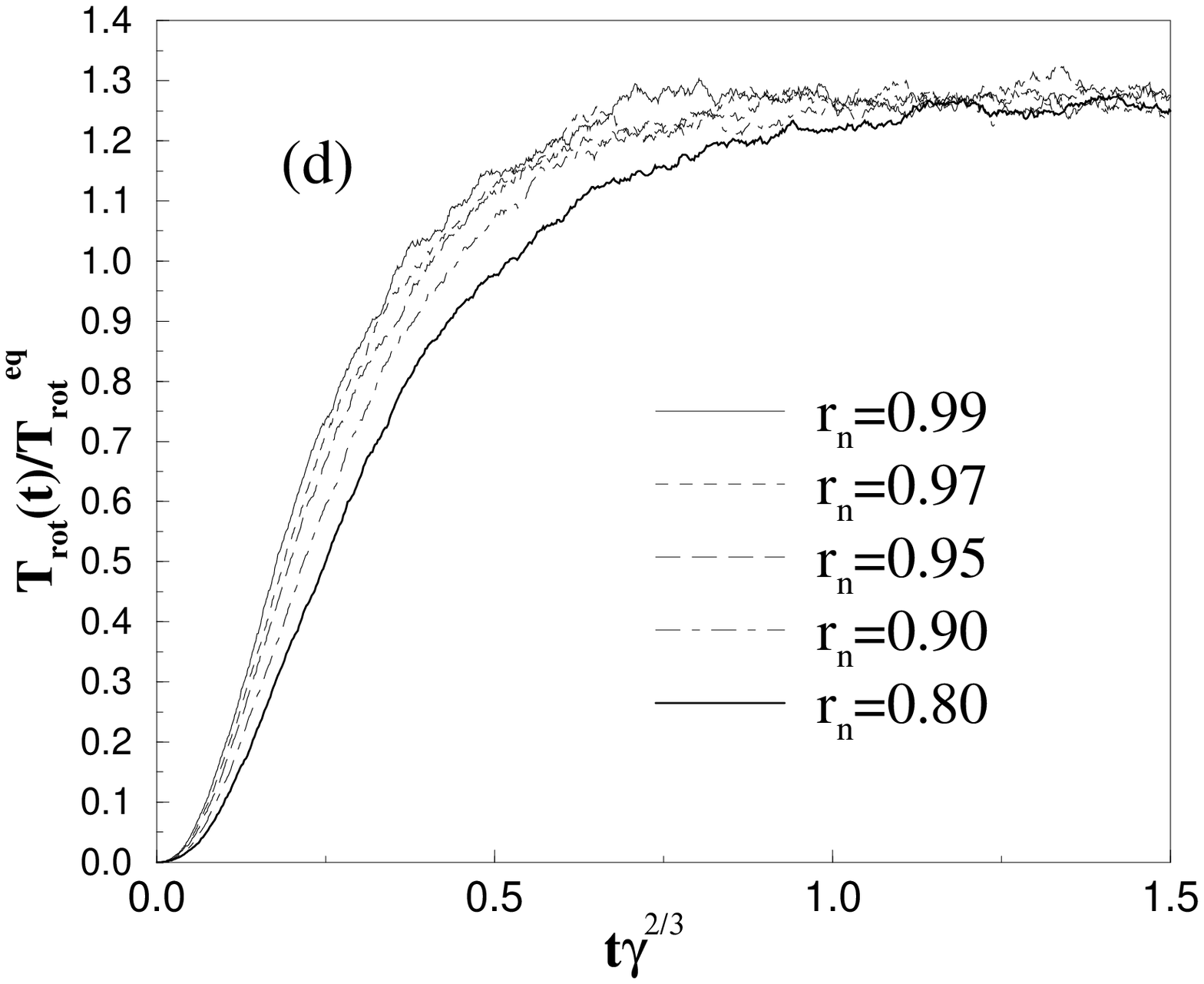,height=4.5cm,width=5.7cm}
\end{center}
\vspace{-0.1cm}
\caption{\footnotesize{(a)-(b) Rescaled translational and rotational 
temperatures $T_{tr}(t)/T_{tr}^{eq}$, $T_{rot}(t)/T_{rot}^{eq}$ 
versus the rescaled time $t\gamma^{2/3}$, for $r_n=0.95,r_t^m=0.4$ 
and different $\mu$; (c)-(d) the same as (a)-(b) but for 
$\mu=0.5,r_t^m=0.4$ and different $r_n$.}}
\label{fig3}
\vspace{0.1cm}
\end{figure}
%
For negative $r_t^m$, agreement with MF is observed for 
the translational temperature, while the rotational temperature 
shows deviations from scaling in the transient phase,
although the equilibrium value fits well with MF, 
for $r_t^m \sim -1$. This is due to a failure of the approximation 
$\delta T_{rot}(t) \simeq  R \delta T_{tr}(t)$ for weak coupling 
between rotational and translational degrees of freedom and finite 
time.  In Fig.\ \ref{fig2}a-b
we plot the same quantities as in Fig.\ \ref{fig1}
but for $r_n=0.95$, $\mu=0.5$. We obtain for 
$r_t^m$ near to unity significant deviations from 
mean field theory, while for negative $r_t^m$ the agreement with 
mean field theory is very good. The explanation for 
this results is that for $r_t^m\sim -1$ and high enough $\mu$, 
one has $\langle{r_t}\rangle \sim r_t^m$, and this correspond to 
have $\mu\to\infty$.  In Fig.\ \ref{fig3}a-b we show 
simulation results for $r_n=0.95$ and variable $\mu$ for $r_t^m=0.4$. These 
simulations confirm the previous interpretation. Finally, in Fig.\
\ref{fig3}c-d we show simulation results for $\mu=0.5$, $r_t^m=0.4$ 
and different $r_n$. In this case the data collapse is very 
good although deviations from MF are observed. 
This is a coincidence, since $\langle{r_t}\rangle\simeq0.363\sim r_t^m$. 
To improve the MF theory for the finite $\mu$ case, the effect of a 
random $r_t$ must be fully taken into account. We actually are studying
the possibility to include a collision 
angle dependent $r_t$ in the ensemble average of the 
energy variation due to collisions \cite{huthmann97}. 
The MF theory we obtain seems to be very promising and 
gives a good qualitative agreement with simulations. A paper is 
in preparation with the results of this study \cite{cafiero99} 
and also a three-dimensional analysis is in progress \cite{herbst99}.
~\vspace{-1cm}\\
\begin{ack}
\vskip -0.5cm
We thank H. J. Herrmann for inspiring discussions and acknowledge 
financial support under the European network project FMRXCT980183
and of the German Science Foundation (DFG).
\end{ack}
~\vspace{-1.5cm}\\

\end{document}